\documentclass{iau}

\usepackage{amsmath}
\usepackage{graphicx}
\usepackage{multirow}

\newcommand\farcs{\mbox{$.\!\!^{\prime\prime}$}}

\begin{document}

\lefttitle{Cambridge Author}
\righttitle{Proceedings of the International Astronomical Union: \LaTeX\ Guidelines for~authors}

\jnlPage{1}{4}
\jnlDoiYr{2025}
\doival{10.1017/xxxxx}

\aopheadtitle{Proceedings IAU Symposium 396}

\title{Formation of Substructure in Luminous Submillimeter Galaxies (FOSSILS): Initial sample and the discovery of a dusty spiral at cosmic noon}

\author{Ryota Ikeda$^{1,2}$, Daisuke Iono$^{1,2}$, Ken-ichi Tadaki$^{3}$, Andrea Silva$^{2}$, \\ and the FOSSILS Team}
\affiliation{$^{1}$ Department of Astronomy, School of Science, SOKENDAI (The Graduate University for Advanced Studies), 2-21-1 Osawa, Mitaka, Tokyo 181-8588, Japan}
\affiliation{$^{2}$  National Astronomical Observatory of Japan, 2-21-1 Osawa, Mitaka, Tokyo 181-8588, Japan}
\affiliation{$^{3}$  Faculty of Engineering, Hokkai-Gakuen University, Toyohira-ku, Sapporo 062-8605, Japan}

\begin{abstract}
High-resolution far-infrared (FIR) observation of submillimeter galaxies (SMGs) is an effective approach to study the formation of substructures in the early epoch of massive galaxies. We present the 870\,$\mu$m continuum images resolved down to sub-kpc scales for 12 SMGs taken by Atacama Large Millimeter/Submillimeter Array (ALMA) as an initial sample of the FOSSILS Survey.  We discovered a wide variety of morphological properties, including a two-arm spiral galaxy at $z=2.5$ possibly induced by a tidal interaction with a minor companion. Nonetheless, about half of the sample exhibit a compact and circular morphology, which is reminiscent of optical morphology of compact quiescent galaxies. Future studies with larger sample and combination of the rest-frame optical images taken by James Webb Space Telescope will shed light on the various evolutionary track of SMGs.
\end{abstract}

\maketitle

\section{Introduction}

How did massive galaxies in the present Universe acquire most of their stellar mass? According to stellar population synthesis of nearby massive elliptical galaxies, it has been suggested that more than half of the old stellar populations were formed by an intense starburst occurred at $z\gtrsim2$ in mass (e.g., \citealp{2010MNRAS.404.1775T}; \citealp{2015MNRAS.448.3484M}). The most likely progenitors of such massive elliptical galaxies are so-called submillimeter galaxies (SMGs; \citealp{2002PhR...369..111B}) at high redshift\footnote{The median redshift of SMGs ($S_{\rm 870\mu m}>1$\,mJy) in the AS2UDS sample is reported to be $z = 2.61$ (\citealp{2019MNRAS.487.4648S}; \citealp{2020MNRAS.494.3828D})}, which can have star formation rates exceeding $1000\,{\rm M}_{\odot}$/yr, consistent with the above formation scenario. 

The unanswered question is how such an intense starburst was triggered in the early Universe. Local analogues of SMGs are ultra-luminous infrared galaxies (ULIRGs), which have comparable infrared (IR) luminosities of $L_{\rm IR}>10^{12}L_{\odot}$. While most nearby ULIRGs are believed to be the aftermath of major mergers, emerging evidence of differing properties between SMGs and local ULIRGs, such as CO sizes and IR surface densities (\citealp{2009ApJ...695..1537I}; \citealp{2011ApJ...726...93R}), implies that major mergers are not the sole mechanism responsible for the extreme starbursts seen in SMGs (\citealp{2013MNRAS.428.2529H}; \citealp{2019MNRAS.488.2440M}).

Formation Of SubStructure In Luminous Submillimeter galaxies (FOSSILS, PI: D. Iono) is an ALMA high-resolution 870\,$\mu$m continuum survey of SMGs, aiming at characterizing the substructures associated with dusty star formation activity. Recently, \cite{2025arXiv251018006I} reported a case study of three luminous SMGs in the COSMOS field combining high-resolution ALMA and James Webb Space Telescope (JWST)/NIRCam imaging. They argued that the physical origin of starbursts are heterogeneous (major merger remnant, gravitationally instability provoked by cold gas inflow, and possible on-going minor mergers), and constructing a statistical sample would be an important next step for further understanding the prevalence of SMGs with different origins. In this proceeding, we present the initial results for the 12 FOSSILS sources and highlight several SMGs with distinctive morphologies. The complete sample, including ALMA archival data, will be presented in our forthcoming work.

\begin{figure}[t!]
  \center
  \includegraphics[width=0.85\linewidth]{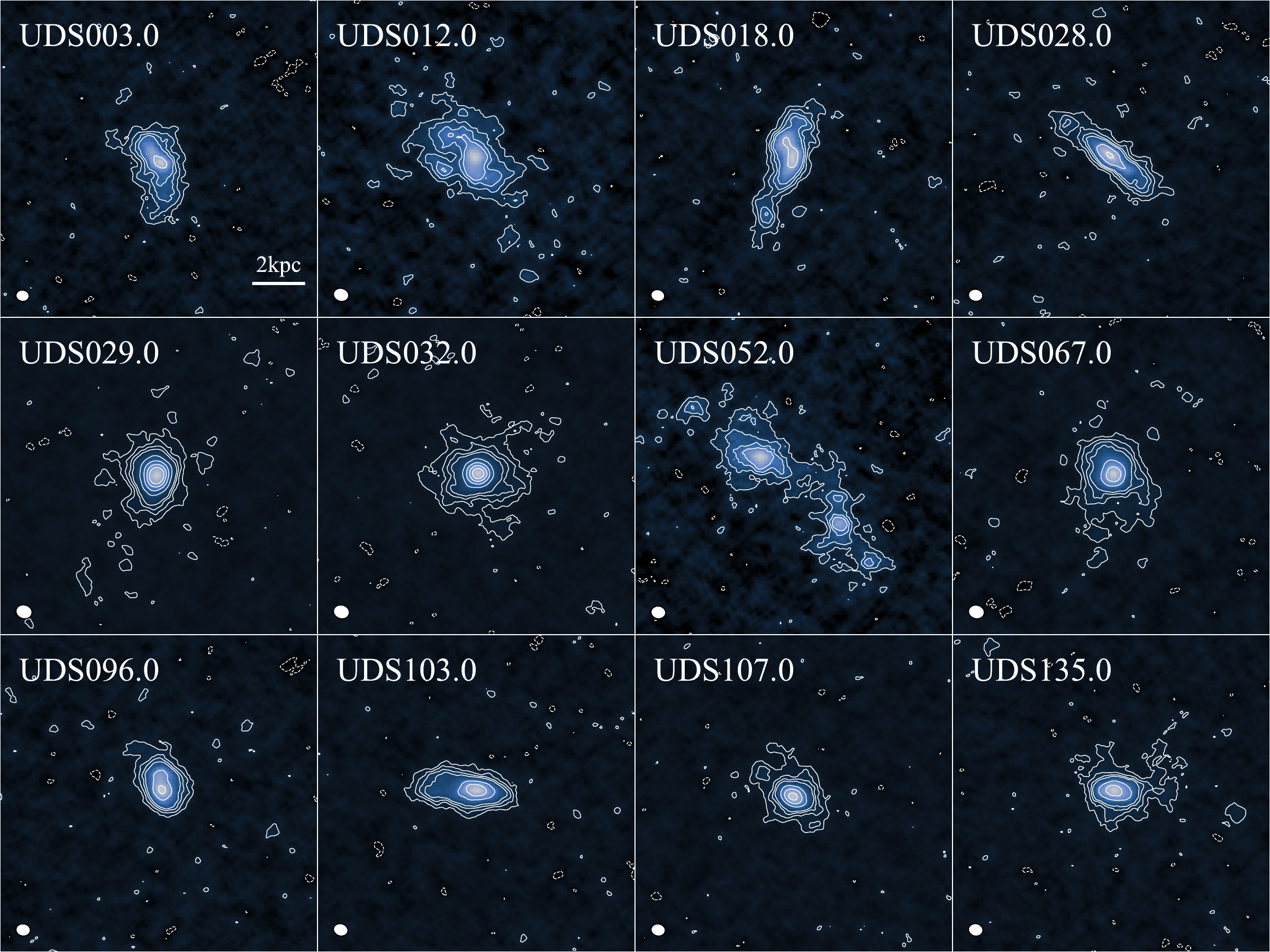}
\caption{Postage stamp images of 870\,$\mu$m continuum emission in the 12 FOSSILS sample. Each panel covers $12$\,kpc$\times12$\,kpc area, scaled according to the redshift of each source. The beam shape is shown in the lower-left corner of each panel. The contour levels are [2.5, 5, 7.5, 10, 20, 30, 40, 50]$\sigma$. }  \label{fig:Figure1}
\end{figure}

\section{Sample and Observations}

The main targets of the FOSSILS Survey are drawn from the ALMA/SCUBA2 UDS (AS2UDS) sample \citep{2019MNRAS.487.4648S}, which are 707 SMGs observed in $0\farcs2$-resolution using ALMA Band 7. Among the 30 SMGs selected based on a flux threshold of $S_{870\mu {\rm m}}>7.8$\,mJy for high-resolution follow-up, only 12 SMGs were observed in June and September 2023 as part of the ALMA Cycle 9 program (\#2022.1.00764.S). The baseline ranges 265-2170 m, corresponding to $uv=$\,300-2500\,k$\lambda$ in spatial frequency.  After calibrating the visibility data using Common Astronomy Software Application package ({\tt CASA}; \citealp{2022PASP..134k4501C}), we combine our data with the visibility data taken in the previous ALMA programs to supplement the short {\it uv} coverage. We applied the CLEAN algorithm to produce the final synthesized image, which was convolved with a 2D Gaussian. We used the natural weighting (robust parameter $R=2.0$) with the CLEAN threshold of $1.5\sigma$ \footnote{For characterizing a two-arm spiral structure in UDS.012, we also applied the Briggs weighting with the robust parameter $R=0.5$. The resulting synthesized beam size is $0\farcs054\times0\farcs044$ (Figure\,\ref{fig:Figure2}).}. The synthesized beam size is approximately $0\farcs075\times0\farcs062$ with a root mean square noise level of $1\sigma=33$-$38$\,$\mu$Jy/beam.

\section{Results}

\begin{figure}[t!]
  \center
  \includegraphics[width=\linewidth]{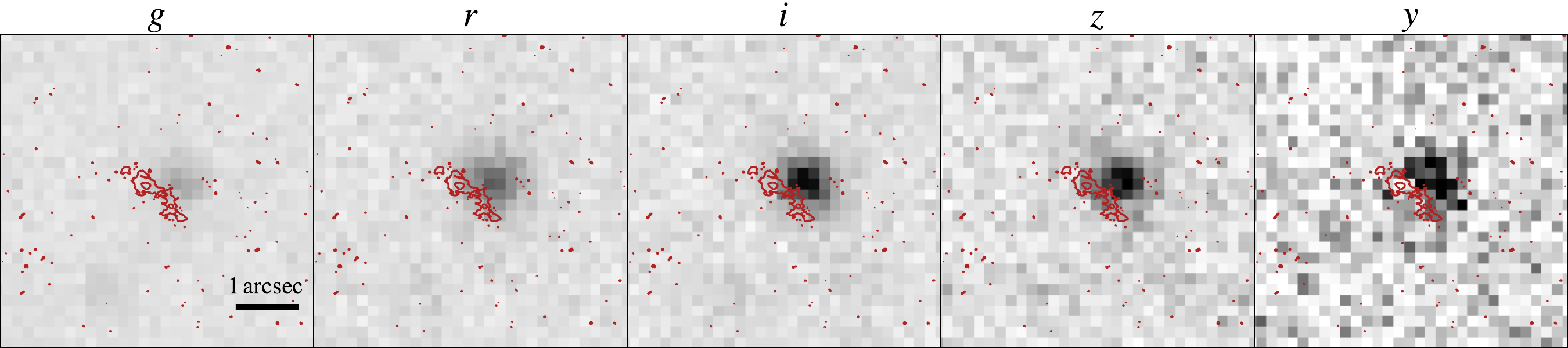}
\caption{Cutouts from the Subaru/HSC-SSP at the position of UDS.052. The astrometric accuracy of the HSC images is $\sim10-20$\,milliarcsecond \citep{2022PASJ...74..247A}. The 870\,$\mu$m continuum contours (3$\sigma$ and 10$\sigma$) are overlaid.}  \label{fig:Figure2}
\end{figure}

Figure\,\ref{fig:Figure1} shows the sub-kpc resolution 870\,$\mu$m continuum emission of the 12 FOSSILS sample. A wide variety of morphologies can be found among the sample, which is consistent with the literature (e.g., \citealp{2016ApJ...829L..10I}; \citealp{2019ApJ...876..130H}). UDS012.0 exhibits a unique spiral-like structure, with an additional clump located $\sim1$\,kpc southeast of the peak position, which we further investigate in the following subsection. UDS052.0 ($z_{\rm phot}=3.13^{+0.22}_{-0.24}$; \citealp{2020MNRAS.494.3828D}) exhibits the most complicated structure among the sample with at least five distinct peaks detected in $>7.5\sigma$ significance level. To our knowledge, this system has not been observed by optical or NIR space telescopes. The images from the Subaru/Hyper-Suprime-Cam Subaru strategic program (HSC-SSP; \citealp{2022PASJ...74..247A}) is shown in Figure\,\ref{fig:Figure2}. A positional offset of $\sim0\farcs2$, which is larger than the astrometric accuracy of the HSC images \citep{2022PASJ...74..247A}, is seen between FIR and optical sources which indicates that UDS.052 may be a gravitationally lensed system. Although we cannot make a conclusive statement without spectroscopic confirmation, UDS.052 may either be an on-going multiple mergers or a gravitationally lensed SMG(s).

While a detailed morphological analysis is awaited, we note that about half of the sample exhibits concentrated morphology without any substructures such as clumps. This possible dichotomy, between irregular/disky and compact morphologies in 870\,$\mu$m continuum emission, may indicate different origins of their starburst activity. Upcoming JWST observations and the study of resolved gas kinematics will further help to confirm this hypothesis.

\subsection{A discovery of two-arm spiral structure in UDS.012 at $z=2.5$}

\begin{figure}[t!]
  \center
  \includegraphics[width=\linewidth]{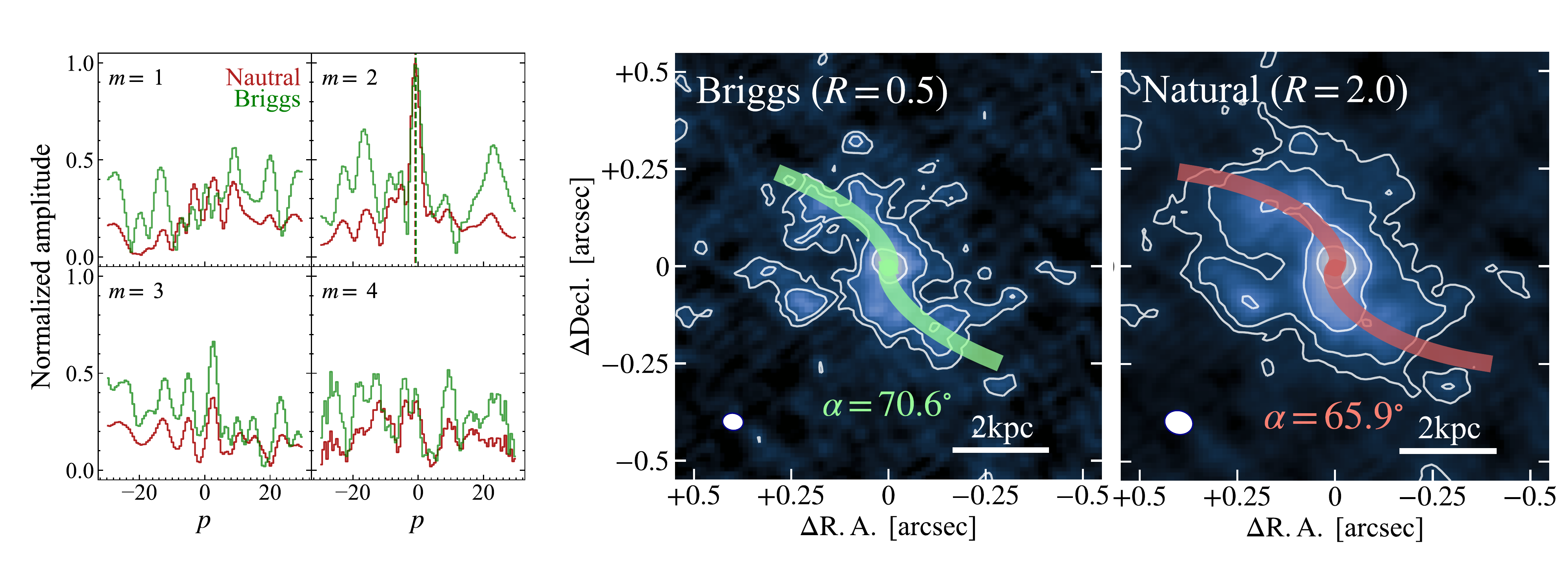}
\caption{A Fourier analysis of UDS.012 at $z=2.52$. Left panel shows a normalized amplitude of four ($m = 1 - 4$) Fourier modes for the images created with Briggs (green) and natural (red) weightings as a function of a dimensionless parameter $p=-m/\tan\alpha$. In either case, the strongest peak is seen in $m = 2$ mode, indicating a two-arm spiral structure with a pitch angle of $\sim65-70^{\circ}$. Middle and right panels show the 870\,$\mu$m images of UDS.012 created with Briggs and natural weightings respectively, with a spiral corresponding to the strongest Fourier component in the $m=2$ mode. The beam shape is shown in the lower-left corner of each panel. The contour levels of [3, 5, 10, 15, 20]\,$\sigma$ are shown.}   \label{fig:Figure3}
\end{figure}

We performed the Fourier decomposition analysis to UDS.012, which revealed a two-arm spiral feature when observed in sub-kpc resolution. We follow the method described in \cite{2021Sci...372.1201T}.  Essentially, this method extracts the strongest Fourier component from the images across different modes ($m$, which can be considered as the number of arms) and pitch angles ($\alpha$). Here, we explore the logarithmic spiral described as $r = r_{0} \exp[-(m/p)\theta]$ in the polar coordinates, where $p=-m/\tan(\alpha)$. Since the inclination of the stellar disk is unknown, we do not apply any correction for the inclination. Figure\,\ref{fig:Figure3} shows the results of the Fourier decomposition using images created with Briggs and natural weightings. We detect the strongest Fourier component in the $m=2$ mode in both images, with a pitch angle of $\sim65-70^{\circ}$, which is broadly consistent with the structure seen in the 870\,$\mu$m image.

The spectroscopic redshift of UDS.012 is reported as $z=2.52$ in \cite{2021MNRAS.501.3926B} based on the detection of the CO\,$J=3-2$ line, followed by the detection of other CO transition lines (\citealp{2023ApJ...945..128F}; \citealp{2025MNRAS.536.1149T}). The mid-IR spectral energy distribution of UDS.012 could indicate the presence of active galactic nuclei \citep{2025MNRAS.536.1149T}. 

One of the mechanisms that can induce an $m=2$ spiral structure is a tidal interaction with a minor merger \citep{2014PASA...31...35D, 2017MNRAS.468.4189P}. A sub-clump structure is detected $\sim1$\,kpc southeast of the main component with a $>5\sigma$ significance. Since the sub-clump is offset from the spiral feature, this may indicate the presence of a minor satellite galaxy responsible for inducing the two-arm spiral. It should be noted, however, that FIR clumps do not necessarily indicate the presence of satellite galaxies, as exemplified by AzTEC-1 at $z = 4.34$ (\citealp{2018Natur.560..613T}; \citealp{2025arXiv251018006I}).
Overall, UDS.012 demonstrates the power of our high-resolution observational campaign in unveiling unprecedented substructures of dust in unlensed SMGs.

\end{document}